\documentclass[trackchanges, twocolumn]{aastex7}
\usepackage{array}
\usepackage{float}

\newcommand{\msun}{M_{\odot}}

\usepackage{amsmath}
\usepackage{makecell, amsmath, caption}
\usepackage{booktabs}
\usepackage{url}
\usepackage{multirow}

\renewcommand{\arraystretch}{1.5}
\begin{document}

\title{An \textit{XMM-Newton} Analysis of the Supermassive Black Hole Binary Candidate MCG+11--11--032}

\correspondingauthor{Yash A. Gursahani}

\author[orcid=0000-0003-1754-2570]{Yash A. Gursahani}
\affiliation{Department of Astronomy, University of Maryland, College Park, 4296 Stadium Dr, College Park, MD 20742, USA}
\email[show]{yashag [at] umd [dot] edu}  

\author[orcid=0000-0001-5766-4287]{Tingting Liu} 
\affiliation{Department of Physics and Astronomy, Georgia State University, 25 Park Place, Suite 605, Atlanta, GA 30303, USA}
\email{tliu20@gsu.edu}

\author[orcid=0000-0002-7962-5446]{Richard Mushotzky}
\affiliation{Department of Astronomy, University of Maryland, College Park, 4296 Stadium Dr, College Park, MD 20742, USA}
\email{rmushotz@umd.edu}

\author[orcid=0000-0002-1510-4860]{Christopher S. Reynolds}
\affiliation{Department of Astronomy, University of Maryland, College Park, 4296 Stadium Dr, College Park, MD 20742, USA}
\affiliation{Joint Space-Science Institute, College Park, MD 20742, USA}
\email{creynold@umd.edu}

\author[orcid=0000-0002-7998-9581]{Michael Koss}
\affiliation{Eureka Scientific, 2452 Delmer Street Suite 100, Oakland, CA 94602, USA}
\email{michaeljkoss@gmail.com}

\begin{abstract}
We investigate the possibility of a binary supermassive black hole system at the center of MCG+11--11--032, a local (z = 0.036) Seyfert 2 galaxy. Prior work with stacked \textit{Swift}/XRT spectra suggested the presence of two Fe K$\alpha$ lines (at 6.16 keV and 6.56 keV) with 2$\sigma$ confidence. This could be consistent with a prediction of several hydrodynamical models, in which each black hole hosts a mini-disk and contributes one Doppler-shifted Fe K$\alpha$ line to the total spectrum. Another study using a single exposure from \textit{Chandra}/ACIS did not find evidence for a double line. Here, we conduct follow-up with two epochs of \textit{XMM-Newton}/EPIC data spaced $\sim$6 months apart. After fitting our spectra with models from the previous two studies, we do not find evidence for a double iron line in either observation. Our best-fit model yields $\Gamma = 1.63^{+0.20}_{-0.21}$ and $N_\text{H}/10^{22} \text{ cm}^{-2} = 17.9^{+2.7}_{-2.4}$ for the first epoch, and $\Gamma = 1.46^{+0.22}_{-0.24}$ and $N_\text{H}/10^{22} \text{ cm}^{-2} = 17.1^{+2.7}_{-2.4}$ for the second. We compare our spectral parameters with those derived in past work on this source, finding broad agreement with prior datasets. Lastly, we discuss the properties of MCG+11--11--032 alongside samples of Seyfert 2 galaxies from the literature, finding that it is consistent with this population and the single AGN scenario. 
\end{abstract}




\section{Introduction} \label{sec:intro}
At the heart of every active galactic nucleus (AGN) lies a supermassive black hole (SMBH), onto which the accretion of gas releases radiation across a broad range of wavelengths (\citealt{salpeter_accretion_1964}, \citealt{kormendy_inward_1995}). Assuming a classical, optically thick $\alpha$-disk \citep{shakura_black_1973}, the observed radiation from a SMBH accretion disk with $M_{\text{BH}} = 10^6-10^9 \msun$ peaks in the optical or ultraviolet bands. These photons are Comptonized by a hot electron population within a few gravitational radii of the black hole (\citealt{bisnovatyi-kogan_disk_1977}, \citealt{zdziarski_physical_1994}), resulting in a power-law X-ray source often called the corona. Photons from the corona may escape directly to the observer or interact with nearby matter. X-rays impinging on the accretion disk or circumnuclear, cold gas produce a so-called ``reflection spectrum," on which the reflecting material leaves an imprint via atomic absorption edges or fluorescent emission lines (\citealt{guilbert_cold_1988}, \citealt{lightman_effects_1988}). 

A particularly important feature of the reflection spectrum is the iron emission region between 6.4-7.1 keV. The Fe K$\alpha$ line from neutral iron is intrinsically a doublet with energies 6.404 keV and 6.391 keV, each having a natural width of only 2-3 eV \citep{holzer_k12_1997}. They are unresolved with CCD observations and their blended profile lies at 6.4 keV. Its energy and shape can be altered by the ionization state of the gas (\citealt{matt_iron_1993}, \citealt{matt_iron_1996}), the transverse Doppler effect \citep{fabian_x-ray_1989}, inclination of the disk \citep{fabian_x-ray_1989}, gravitational redshift as a function of black hole spin (\citealt{fabian_x-ray_1989}, \citealt{laor_line_1991}, \citealt{fabian_broad_2000}), and chemical abundances \citep{garcia_x-ray_2010}. The theoretical modeling of this line has enabled its use as an observational diagnostic to validate various physical scenarios, including complex absorption along the line of sight, relativistic reflection, and gas ionization (e.g. \citealt{mushotzky_x-ray_1978}, \citealt{nandra_detection_1989}, \citealt{pounds_iron_1989}, \citealt{pounds_x-ray_1990}, \citealt{matsuoka_x-ray_1990}, \citealt{tanaka_gravitationally_1995}).

If we suppose that every galaxy hosts a SMBH, then supermassive black hole binaries (SMBHBs) should be a natural byproduct of galaxy mergers. The two black holes first sink toward the center of the post-merger system due to dynamical friction on $\sim$Gyr timescales (\citealt{begelman_massive_1980}, \citealt{binney_tremaine_gd}). We call these systems dual AGN if both black holes are actively accreting at kiloparsec-scale separations. Hardening, which is the decreasing of the SMBH separation \textit{a}, can proceed to sub-parsec separations due to environmental interactions with gas, stars, or other black holes (\citealt{mayer_rapid_2007}, \citealt{berczik_efficient_2006}, \citealt{gualandris_collisionless_2017}, \citealt{ryu_interactions_2018}). At this stage, we call the system a SMBHB, as the black holes interact with one another gravitationally. Once formed, the binary orbit is predicted to reside within a low-density cavity surrounded by a circumbinary disk. Streams of gas escape into the cavity, feeding ``mini-disks" around each individual black hole. This physical picture is a result of various hydrodynamical models, with some including magnetic effects (MHD), general relativity (GRMHD), and 2D or 3D simulation domains (e.g. \citealt{bowen_relativistic_2017}, \citealt{ryan_minidisks_2017}, \citealt{tang_orbital_2017}, \citealt{bowen_quasi-periodic_2018}, \citealt{dascoli_electromagnetic_2018}). There have been several proposed methods by which one could find accreting sub-parsec SMBHBs, or binary AGN. These include detecting periodically modulated emission (e.g. \citealt{dorazio_reduced_2015}, \citealt{tang_orbital_2017}, \citealt{dorazio_repeated_2018}), X-ray shock signatures from gas streams colliding with mini-disks (\citealt{roedig_observational_2014}, \citealt{farris_characteristic_2015}), or periodic oscillation of binary X-ray spectral features \citep{jovanovic_composite_2014}. Despite having these predictions at our disposal, sub-parsec SMBHBs have proven elusive due to the difficulty of distinguishing their emission from that of single AGN. Confirming their presence remains an observational challenge.

For binary AGN, X-ray reflection could be observable from both mini-disks individually. This could manifest as two distinct Fe K$\alpha$ lines that shift in energy space over a single orbit due to the orbital motion of the binary. The outlook for spectrally resolving the lines improves at closer separations ($\sim0.01$ pc), lower binary inclinations ($i \lesssim 60^{\circ}$) and mass ratios approaching unity (\citealt{sesana_multimessenger_2012}, \citealt{jovanovic_composite_2014}). As an example, if we approximate $v/c \sim \Delta E/E$, then the maximum energy shift of the line given a line-of-sight orbital velocity $\sim10^4$ km s$^{-1}$ would be $\Delta E_{\text{max}} \sim 0.2$ keV. In this case, the two lines would complete one cycle in just a few years for an equal-mass binary with $M_{\text{tot}} \sim 10^9 M_{\odot}$ in a circular orbit.

MCG+11--11--032 (hereafter referred to simply as MCG+11), is a local (z = 0.036) Seyfert 2 with log($M_{\text{BH}}/\msun$) = $8.7 \pm 0.3$ \citep{koss_bat_2017}. A pair of [OIII] emission lines in its SDSS-DR7 spectrum previously earned MCG+11 consideration as a dual ($a\sim$ kpc) AGN candidate \citep{wang_active_2009}, assuming the velocity offset in the two AGN narrow line regions manifests as velocity shifts in narrow emission lines. Follow-up with ground-based slit spectroscopy yielded a similar result \citep{comerford_kiloparsec-scale_2012}.  \citet{severgnini_swift_2018} used the MCG+11 optical spectrum from SDSS-DR13 \citep{albareti_13th_2017} to model several lines with two Gaussian components, including [OIII], H$\alpha$, [NII], and [SII]. The results were broadly consistent with \citet{wang_active_2009} and \citet{comerford_kiloparsec-scale_2012}, with the two Gaussians showing velocity offsets of $\sim$250-300 km s$^{-1}$ in all cases. MCG+11 is also a hard X-ray source, observed as part of the 70-month \textit{Swift}/BAT all-sky survey (\citealt{baumgartner_70_2013}; \citealt{barthelmy_burst_2005}). \citet{severgnini_swift_2018} used the 123-month BAT lightcurve to study long-term variability of the source, which seemed consistent with a putative periodicity of $P=25$ months. \citet{serafinelli_unveiling_2020} examined the BAT lightcurve, accounting for possible colored noise. They found a peak in the power spectral density (PSD) corresponding to a period of $26^{+4}_{-3}$ months at $2.6\sigma$ significance, consistent with \citet{severgnini_swift_2018}. \citet{liu_bat_2020} conducted an independent search, fitting the power spectra of hundreds of sources and finding that MCG+11 was red noise-dominated without any prominent periodic signal. The different analysis techniques may account for the inconclusive results, highlighting the difficulty of using periodicity to find binary AGN. Complementary methods, including spectroscopy, are necessary tools in the search for SMBHBs. For example, \citet{severgnini_swift_2018} used archival and targeted \textit{Swift}/XRT \citep{burrows_swift_2005} observations to examine the Fe K$\alpha$ emission. Instead of one line at 6.4 keV, they found tentative evidence for two lines: one at 6.16 keV and another at 6.56 keV with 2$\sigma$ confidence. If confirmed, this could provide strong evidence for a sub-parsec SMBHB, with each line Doppler shifted away from its rest-frame value due to the orbital motion of mini-disks. A double iron line combined with a 25-month variability timescale and the aforementioned total black hole mass would suggest an orbital speed of $\sim$0.06$c$, or $1.8 \times 10^4$ km s$^{-1}$.

The result presented by \citet{severgnini_swift_2018} relied on stacking several epochs of XRT spectra. Assuming the presence of the SMBHB and a $\sim2$ yr orbital period, the two Fe K$\alpha$ lines would have shifted back and forth in energy space several times over the course of those observations ($\sim7$ years). Recently, \citet{foord_chandra_2025} used \textit{Chandra}/ACIS to take a single, deep exposure of MCG+11 in March 2023. The spectrum shows no evidence for a second Fe K$\alpha$ line in the energy range expected from \citet{severgnini_swift_2018}. However, to confirm the double black hole scenario, it is not only necessary to use sensitive telescopes, but also to observe the source over multiple epochs and search for spectroscopic line shifts consistent with the binary orbital motion. Therefore, an observing strategy that could more robustly test the double line model includes two exposures taken a few months apart with a similarly sensitive telescope. If the period of the binary were 25 months, the positions of the two Fe K$\alpha$ lines should appear differently in each spectrum. If the single AGN model is preferred, we expect one line with consistent energies in both spectra. This is the approach adopted in this work; we use the \textit{XMM-Newton}/EPIC instrument to observe MCG+11 roughly 6 months apart and model its spectrum, searching for changes in the iron emission on this timescale. 


This paper is organized as follows: Section \ref{sec:obs} discusses the observations we use in this work along with the details of the data reduction. Section \ref{sec:analysis} contains our spectral analysis, results, and comparisons to previous work. Lastly, we summarize and conclude in Section \ref{sec:conclusion}. In this work we assume a $\Lambda$CDM cosmology with $H_0 = 70.0$ km s$^{-1}$ Mpc$^{-1}$, $\Omega_m = 0.27$, and $\Omega_\Lambda = 0.73$.

\section{Observations and Data Reduction} \label{sec:obs}

\begin{table*}[]
    \centering
    \begin{tabular}{c|c|c|c|c|c}
        OBSID & Obs. Date & Detector & $t_{\text{exp, tot}}$ (ks) & $t_{\text{exp, filt}}$ (ks) & Source Count Rate (cts s$^{-1}$) \\
        \hline 
        0921640101 & 2023-11-14 & EPIC-pn & 48.7 & 27.3 & $0.115\pm0.002$ \\
        &  & EPIC-MOS1 & 50.6 & 36.7 & $0.032\pm0.001$ \\
        &  & EPIC-MOS2 & 50.6 & 37.0 & $0.033\pm0.001$ \\
        0921640201 & 2024-05-01 & EPIC-pn & 51.4 & 22.6 & $0.130\pm0.003$ \\
        &  & EPIC-MOS1 & 42.1 & 29.9 & $0.036\pm0.001$ \\
        &  & EPIC-MOS2 & 56.3 & 36.2 & $0.037\pm0.001$ \\
    \end{tabular}
    \caption{Summary of observational data used in this work. $t_{\text{exp, tot}}$ is the total exposure time recorded for the detector while $t_{\text{exp, filt}}$ is the filtered time after applying GTIs. The reported count rate for MCG+11 is also after filtering.}
    \label{tab:observations}
\end{table*}

Data from \textit{XMM-Newton} were obtained via targeted observations of MCG+11--11--032 ($08^h55^m12.58^s$, $+64^{\circ}23'45.42^"$; J2000) in AO-22 (PI: T. Liu, OBSIDs: 0921640101 and 0921640201). Two exposures were carried out, the first on 14 November 2023 and the second on 1 May 2024. See Table \ref{tab:observations} for details of each observation. The data from EPIC-MOS \citep{turner_european_2001} and EPIC-pn \citep{struder_european_2001} were reduced using the \textit{XMM} Science Analysis Software (\texttt{SAS}) V21.0.0 \citep{gabriel_xmm-newton_2004}. We will now describe the data reduction procedure for each detector, since the procedures vary slightly.


To filter for particle flaring, we examined the full-FOV, hard X-ray lightcurves for each detector. Following the standard procedure outlined in the SAS User Guide \citep{de_la_calle_users_2023}, we identified threshold count rates by eye below which we consider the data to be dominated by astrophysical X-rays. For the pn, we carried out the filtering in the 10-12 keV band, setting thresholds of 0.26 cts/s for the first observation and 0.57 cts/s for the second observation. The good-time intervals (GTIs) total 27 ks and 23 ks for the first and second epochs, respectively. We note that these cuts result in a loss of $\sim50\%$ of the exposure time for the pn, which is larger than the average loss due to flaring ($\sim20-30\%$). Next, we chose source and background extraction regions. The pn source region is a circle of radius of 30" while the background is a circle with radius 50". There are electronics below the pn CCD that produce K$\alpha$ fluorescent lines from Ni (7.5 keV), Cu (8.0 keV), and Zn (8.6 keV) in predictable locations when exposed to X-rays \citep{freyberg_xmm-newton_2006}. These lines are in the hard X-ray band and have energies reasonably close to the Fe K$\alpha$ line. We make sure to extract our background from the same region where the source lies. In this way, we avoid artifacts when subtracting the background in our analysis. Since we are trying to detect and characterize lines as a key diagnostic of SMBHBs, this prevents any possible contamination of our spectra.

We applied our flaring thresholds for the MOS detectors above 10 keV, again following the standard SAS procedure. For MOS1, we removed data above 0.07 cts/s and 0.20 cts/s for the first and second observations, respectively. For MOS2, these values were 0.10 cts/s and 0.50 cts/s. The GTIs total 37 ks and 30 ks for MOS1, while the total is 37 ks and 36 ks for MOS2. The filtering reduces the exposure time by $\sim30\%$ for the MOS detectors. For both MOS detectors, we use a source extraction region of radius 25" with an annular region of outer radius 60" centered on the source for the background. 

The spectra from each observation were binned with the \texttt{ftgrouppha} task using the optimal binning scheme \citep{kaastra_optimal_2016} and the additional requirement of a minimum of 30 counts per bin. We produced the photon redistribution matrix file (RMF) and ancillary response file (ARF) for all detectors within \texttt{SAS} using the tasks \texttt{rmfgen} and \texttt{arfgen}, respectively. We analyze the 2-10 keV spectrum from the pn and the 2-9 keV spectra from MOS1 and MOS2, due to low photon counts at high and low energies.

\section{Analysis and Results} \label{sec:analysis}
Here, we will describe the models used in this work and their resulting fit parameters. We carry out identical analyses for both epochs, fitting MOS and pn spectra jointly in \textsc{xspec} V.12.14.1 \citep{arnaud_xspec_1996}. Elemental abundances in this work are taken from \citet{anders_abundances_1989} and cross sections are from \citet{verner_atomic_1996}. All quoted energies are in the rest frame of the source. 

\subsection{Spectral Fitting and Testing the Double Iron Line Hypothesis With \textit{XMM}} \label{subsec:iron_line}
\begin{figure*}
    \centering
    \includegraphics[width=0.7\textwidth]{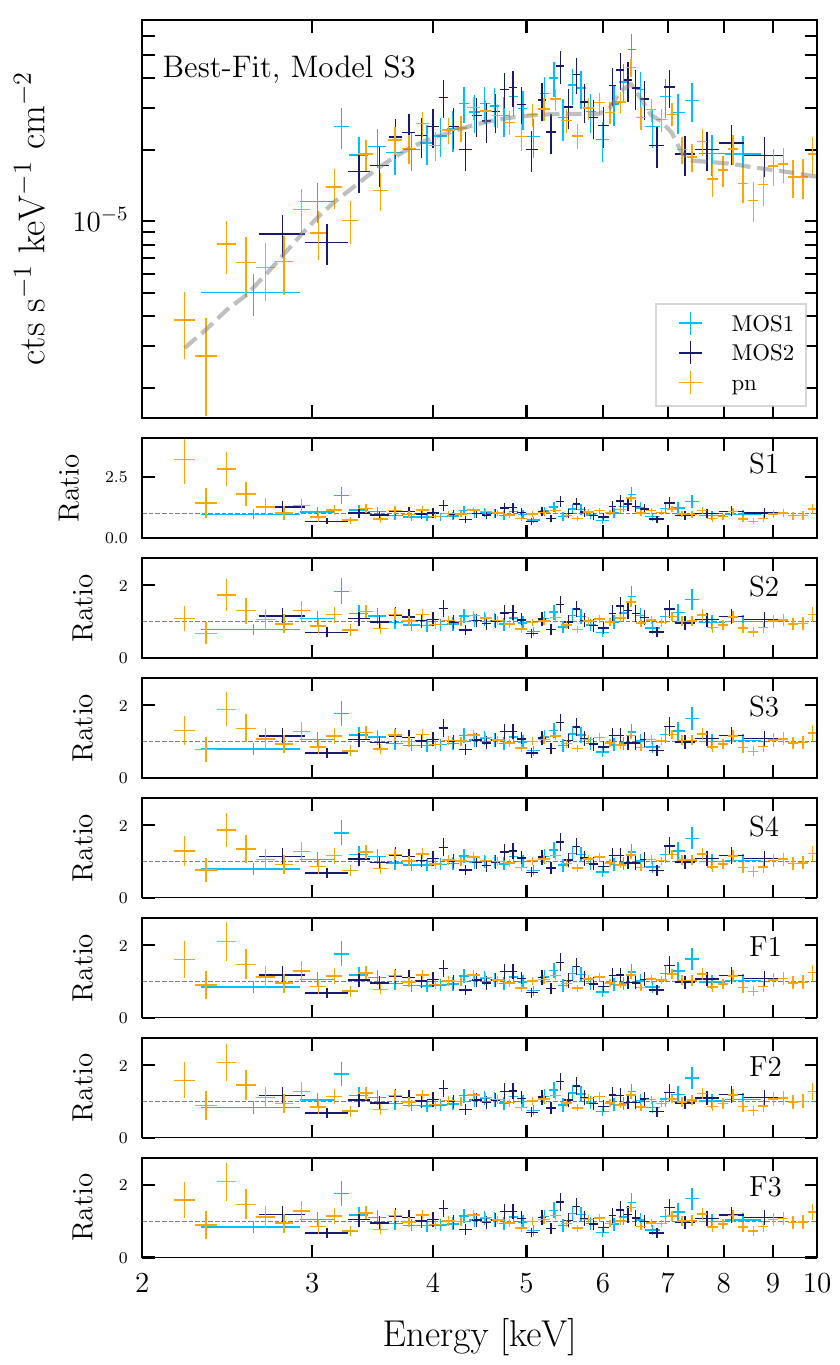}
    \caption{Rest-frame spectrum and data-to-model ratios from the first epoch of \textit{XMM}/EPIC observations. We reproduce the models from \citet{severgnini_swift_2018} and \citet{foord_chandra_2025}, fitting them to the data. The main panel shows the spectrum and the best-fit model, S3, indicated with the dashed line. The panels below this show the ratio of the data to each model from the previous two papers, with the dashed line at a ratio of 1. Note that the ratio plot for model S1 has a different vertical scale than the others; we highlight that this is a poor fit due to large residuals. Each panel is labeled with the model name, for which the parameters can be found in Table \ref{tab:pn_obs1}. Model S1 consists of an absorbed power-law. Model S2 adds a slab reflection component. Model S3 uses one Gaussian to fit the Fe K emission region. Finally, model S4 tests for the presence of a second iron line. Model F1 consists of an absorbed power-law with slab reflection and a Gaussian iron line at $\sim$6.4 keV, similar to model S3. Model F2 forces another Gaussian to fit between 6-7 keV. Model F3 implements the two best-fit line energies from \citet{severgnini_swift_2018}, 6.16 keV and 6.56 keV.}
    \label{fig:sevfoord_obs_1}
\end{figure*}

\begin{figure*}
    \centering
    \includegraphics[width=0.7\textwidth]{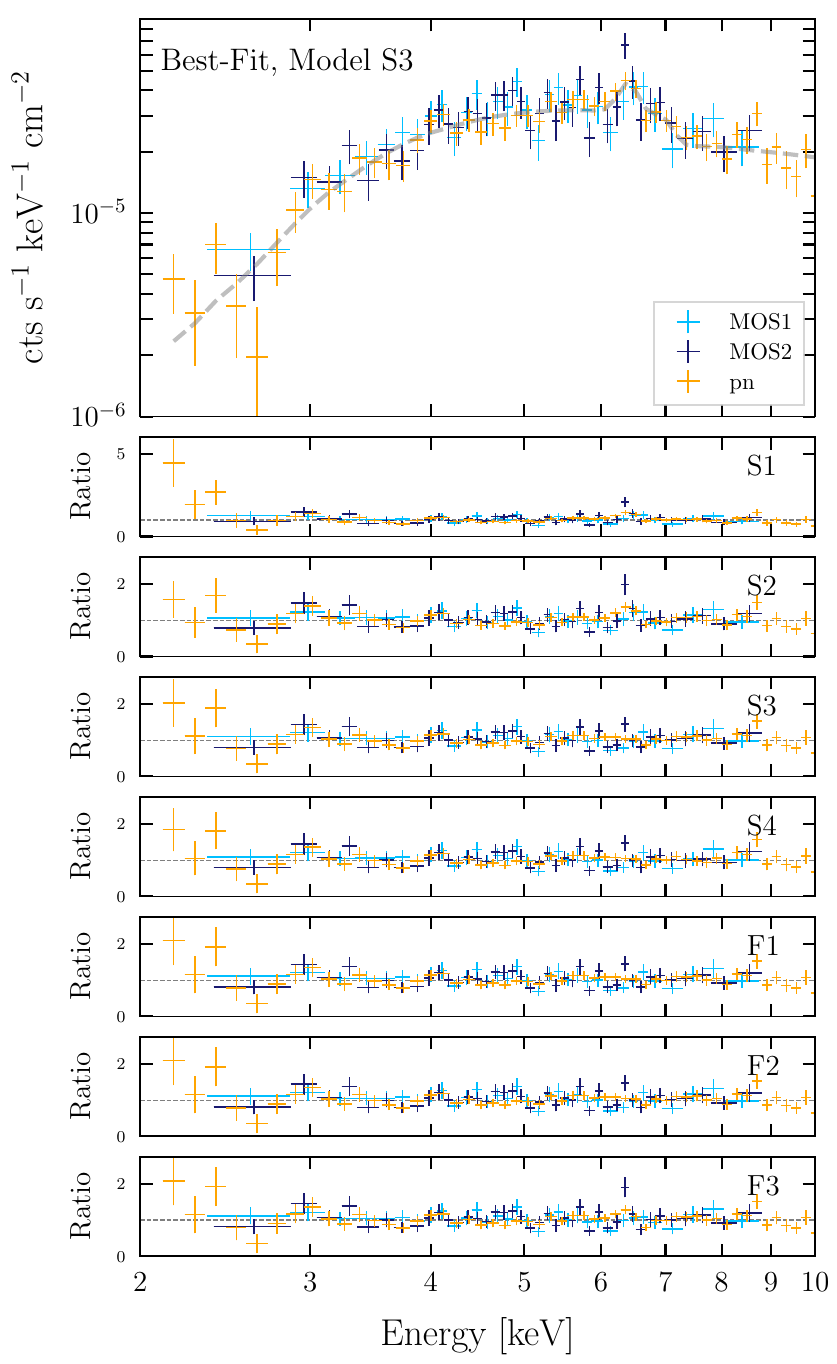}
    \caption{Same as Figure \ref{fig:sevfoord_obs_1} for the second epoch of our \textit{XMM}/EPIC data.}
    \label{fig:sevfoord_obs_2}
\end{figure*}

In order to test for the presence and significance of the putative double Fe K$\alpha$ line around the rest-frame value of 6.4 keV, we begin by fitting \textit{XMM}/EPIC data with the four models from \citet{severgnini_swift_2018}. We name these models S1, S2, S3, and S4. Model S1 consists of a simple absorbed power-law: \texttt{tbabs*ztbabs*zpowerlw}, which describes an incident power-law from a Comptonizing source such as a black hole corona being absorbed by the ISM intrinsic to the source at redshift $z$ and then by the Milky Way \citep{wilms_absorption_2000}. We fix the value of the Galactic column density at $N_{\text{H, MW}} = 4.7 \times 10^{20}$ cm$^{-2}$ in all our fits \citep{kalberla_leidenargentinebonn_2005}. The best fit for this model has $\chi^2$/dof = 166.46/119 for the first observation and 147.02/120 for the second observation. Thus, model S1 is insufficient to describe the data, as evidenced by the S1 ratio panel in Figures \ref{fig:sevfoord_obs_1} and \ref{fig:sevfoord_obs_2}. There are large residuals around 6.4 keV. Additionally, we note an excess below 3 keV. Possible explanations for this are the narrow-line region \citep{bianchi_evidence_2019}, scattering of the primary X-ray emission \citep{ueda_suzaku_2007}, or star-formation \citep{ranalli_2-10_2003}. \citet{severgnini_swift_2018} and \citep{foord_chandra_2025} model this excess as reprocessing by circumnuclear material (e.g. \citealt{lamer_rxte_2000}, \citealt{alexander_nustar_2013}, \citealt{del_moro_nustar_2017}). We adopt their choice of the \texttt{pexrav} model \citep{magdziarz_angle-dependent_1995} to ensure a fair comparison between datasets. 

Therefore, model S2 is \texttt{tbabs*(ztbabs*zpowerlw + pexrav)}. Note that we use \texttt{pexrav} only to capture the reflected emission; this means that we force the $rel_{refl}$ parameter to be negative while tying both the power-law index $\Gamma$ and the normalizations of \texttt{pexrav} and \texttt{zpowerlw} together. $rel_{refl}$ is allowed to fit freely between 0 and -10. The inclination is not well-constrained, but the fit prefers higher values for cos\,$(i)$. Since the fits are insensitive to this angle, we fix cos\,$(i) = 0.95$, which produces physically reasonable values for $rel_{refl}$ as opposed to lower cos\,$(i)$. All abundances are fixed to their default, solar values. For only one additional parameter, the improvement in the fit is $\Delta\chi^2 = -19.20$ for the first observation and $\Delta\chi^2 = -10.72$ for the second, translating to significance levels of 4.4$\sigma$ and 3.3$\sigma$, respectively. The 2-3 keV residuals in the S2 ratio panels of Figures \ref{fig:sevfoord_obs_1} and \ref{fig:sevfoord_obs_2} are drastically reduced, and so we conclude that a reflection model is statistically required to describe the data. 

\begin{figure*}
    \centering
    \includegraphics[width=0.7\textwidth]{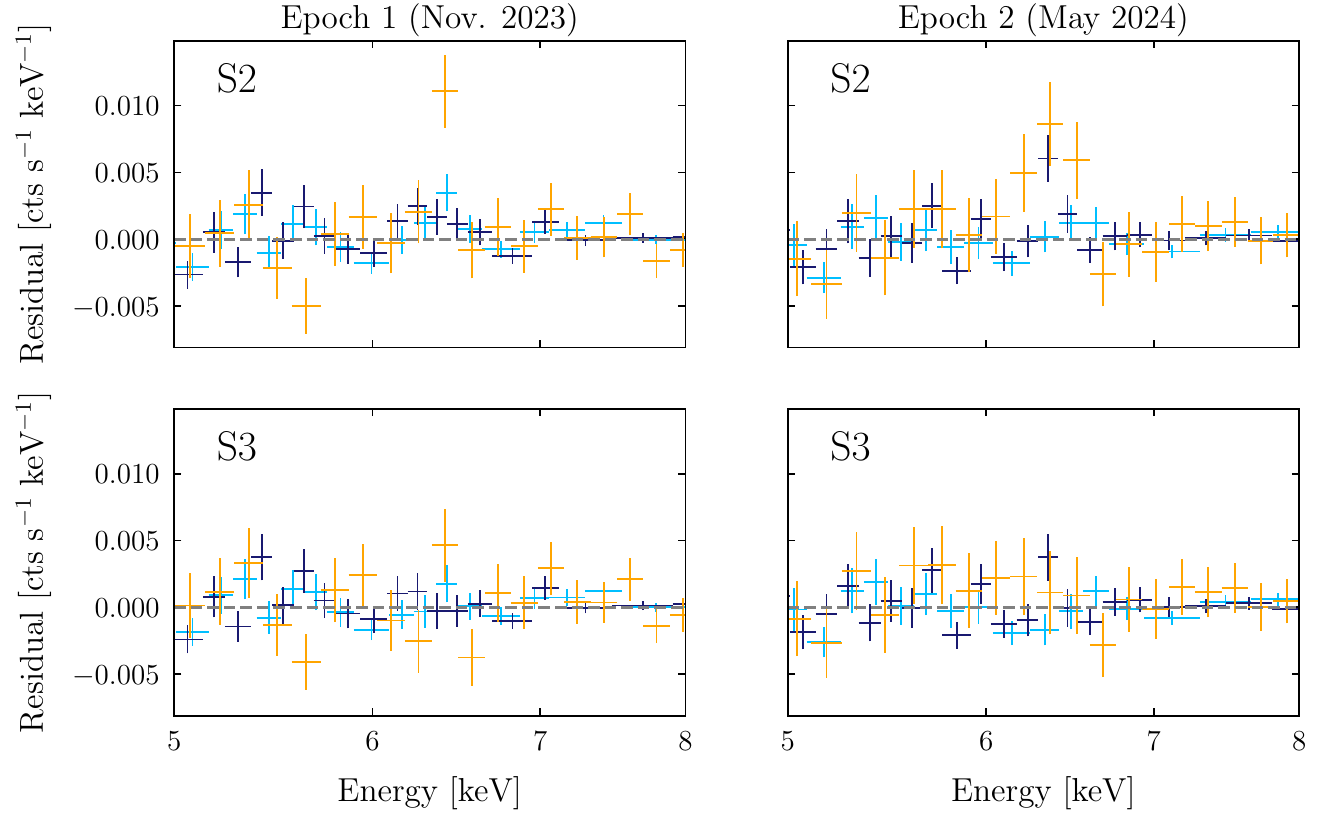}
    \caption{5-8 keV residuals between models S2 and S3 for the first epoch in November 2023 (left column) and second epoch in May 2024 (right column). As in Figures \ref{fig:sevfoord_obs_1} and \ref{fig:sevfoord_obs_2}, we show data from MOS1 (light blue), MOS2 (dark blue), and the pn (orange). Note the difference in residuals between the models at 6.4 keV, indicating the addition of a Gaussian Fe K$\alpha$ emission line in model S3.}
    \label{fig:s2_s3}
\end{figure*}

To address the notable residual around 6.4 keV, model S3 adds a Gaussian emission line to account for a single Fe K$\alpha$ component: \texttt{tbabs*(ztbabs*zpowerlw + pexrav + zgauss)}. Often, the line widths ($\sigma_\text{line}$) are not well-constrained and the fit prefers extremely small values beyond the resolution of the EPIC detectors. In this case, we fix $\sigma_\text{line}$ to its 90\% upper limit, which falls between 85 and 125 eV. If \textsc{xspec} fails to place an upper limit on the width, we fix it to 100 eV, which is a reasonable width for a broad Fe K$\alpha$ line (e.g. \citealt{nandra_xmm-newton_2007}). The addition of one Gaussian line results in a reduction in the fit statistic for both epochs: $\Delta\chi^2 = -23.05$ for the first and $\Delta\chi^2 = -20.78$ for the second. The Fe K$\alpha$ line is therefore statistically significant at a $>4\sigma$ level in both observations. Note the reduction of the residual at 6.4 keV between the `S2' and `S3' panels of Figures \ref{fig:sevfoord_obs_1} and \ref{fig:sevfoord_obs_2}. Figure \ref{fig:s2_s3} shows the comparison of S2 and S3 residuals plotted in count rate units, from which this improvement may be more apparent. $\chi^2$/dof $\sim1$ now for both epochs, indicating an acceptable fit to the spectrum of MCG+11. 

Yet, there are some small features that are not fully explained by our model. We would like to assess whether they arise from other physical means or can be explained simply as noise. Model S4 adds another Gaussian to probe for a second emission line in the 4-8 keV range: \texttt{tbabs*(ztbabs*zpowerlw + pexrav + zgauss + zgauss)}. For the first epoch, this line fits best at an energy of 4.54 keV. Similarly, the energy we derive for the second epoch is roughly 7.54 keV. We note that both these line energies are quite poorly constrained. The improvements in the fit statistics for the first and second epochs are $\Delta\chi^2 = -1.73$ and $\Delta\chi^2 = -1.41$, respectively. These small changes indicate that adding two parameters for a second emission line is not justified for either observation. In addition, for an edge-on system they imply an orbital speed of $\gtrsim0.15c$ or period of $\sim$days, which is inconsistent with the $\sim2$ yr orbit suggested in earlier work. All parameters and statistics for these fits can be found in the first four rows of Table \ref{tab:pn_obs1} for the first observation and Table \ref{tab:pn_obs2} for the second observation. 

The next set of models comes from \citet{foord_chandra_2025}, who studied the same source with \textit{Chandra}. In keeping with our naming convention, our names for these are F1, F2, and F3. Model F1 begins with the best-fit model from \citet{severgnini_swift_2018}, model S3: \texttt{tbabs*(ztbabs*zpowerlw + pexrav + zgauss)}. They make a slight alteration to this model: \citet{foord_chandra_2025} choose to fix $rel_{refl} = -1$ rather than allow it to vary. In doing so, they assume that the reflected emission is equivalent to that from a power-law source impinging on a semi-infinite slab that covers half of the sky from the source's point of view. The Gaussian line is initialized at 6.16 keV, which was the best-fit found by \citet{severgnini_swift_2018} for model S3. It is allowed to vary only between 6-7 keV, which is a tighter constraint than the previous set of models. This is justified, given that shifting the neutral Fe K$\alpha$ line from 6.4 keV to 7 keV would translate to an extremely high, maximum orbital speed of $v\sim0.1c$ for one of the SMBHs in a binary. $\Gamma$ values are kept within 1-3, which is a range consistent with the broader AGN population (e.g. \citealt{ricci_bat_2017}). For both epochs of our data, this results in a line centered around $\sim6.4$ keV rather than the initial 6.16 keV. The fit is acceptable, but slightly worse than model S3 with $\chi^2$/dof = 125.39/117 for the first observation and $\chi^2$/dof = 115.66/118 for the second.

Model F2 initializes another Gaussian line at 6.56 keV, which was the best-fit found for model S4 by \citet{severgnini_swift_2018}. As with the first Gaussian, we allow it to fit only between 6-7 keV. In the first epoch, the line settles at the upper bound of 7 keV while in the second epoch, it prefers the 6 keV lower bound. In neither case does it appear statistically significant, and the uncertainties on the line energy cannot be estimated by the standard \textsc{xspec} \texttt{error} command. In order to calculate the equivalent width and associated uncertainty for weakly-constrained lines, we fix them to their best-fit energies and refit.

Lastly, model F3 \textit{fixes} the two line energies at 6.16 keV and 6.56 keV. The fit using model F3 has $\Delta\chi^2 = 14.21$ (Obs. 1) and $\Delta\chi^2 = 23.84$ (Obs. 2) compared to model F1, indicating a poorer fit for one additional parameter in Obs. 1 and the same model complexity in Obs. 2. From our analysis, the \textit{XMM} data do not require a second line.

It is worth noting here that the models we fit in this work are largely phenomenological and do not represent any one physical scenario. In particular, our use of a Gaussian to describe an emission line does not take into account the true shape of the line nor does it explicitly link it to the \texttt{pexrav} component. We expect an X-ray source with a slab covering half the sky ($rel_{refl} = -1$) to produce an Fe K$\alpha$ equivalent width (EW) of $\sim$150-200 eV \citep{george_x-ray_1991}. Our best-fit values for $rel_{refl}$ and calculated EWs are consistent with this expectation. However, to test a more physically motivated model, we replace \texttt{pexrav + zgauss} with the \texttt{pexmon} model \citep{nandra_xmm-newton_2007}, which self-consistently includes Fe K$\alpha$, Fe K$\beta$, and Ni K$\alpha$ emission lines that result from slab reflection. As before, we fix the inclination to $18.2^{\circ}$ (cos$\, i = 0.95$) to fairly compare between models. The fit statistics with \texttt{pexmon} are slightly better than model S3 and have fewer free parameters. Our best-fit $\Gamma$ and $N_\text{H}$ are generally consistent between the two fits, but $rel_{refl}$ now takes on a value of $\sim0.55$ in the first epoch and $\sim0.45$ in the second. Using formulas in \citet{nandra_xmm-newton_2007}, we arrive at Fe K$\alpha$ EWs of 300-500 eV, which are much higher than that measured in model S3 with a simple Gaussian. Additionally, the reflected component below 3 keV is weakened with \texttt{pexmon}, which results in large residuals for the pn data. Ultimately, we choose not to include this model in our figures or tables in order to directly compare to the results from \citet{severgnini_swift_2018} and \citet{foord_chandra_2025}.



Now, we briefly summarize the models and results from this section. Our best-fit model for both epochs is model S3, which includes an absorbed power-law, slab reflection, and a single Gaussian emission line capturing the iron emission. We conclude, in accordance with \citet{foord_chandra_2025}, that there is insufficient evidence from the X-ray spectra of MCG+11 to support the double Fe K$\alpha$ line model. As we show in Tables \ref{tab:pn_obs1} and \ref{tab:pn_obs2}, the fit parameters from the first and second observations agree with one another.


\begin{table*}
    \centering
    \begin{tabular}{c|c|c|c|c|c|c|c|c|c}
    
         \multirow[t]{3}{4em}{(1) Model} & (2) & (3) & (4) & (5) & (6) & (7) & (8) & (9) & (10) \\
         & $\Gamma$ & $N_{\text{H}}$ & $rel_{refl}$ & $E_\text{line}$ & $\sigma_{\text{line}}$ & EW & $F_\text{2-10}$ ($10^{-12}$ & $L_\text{2-10}$ ($10^{42}$ & $\chi^2$/dof \\
         & & ($10^{22}$ cm$^{-2}$) & & (keV) & (eV) & (eV) & erg s$^{-1}$ cm$^{-2}$) & erg s$^{-1}$) \\
         
         \hline

         S1 & $1.41^{+0.17}_{-0.16}$ & $15.8^{+1.5}_{-1.4}$ & -- & -- & -- & -- & $1.61^{+0.03}_{-0.14}$ & $4.63^{+0.12}_{-0.38}$ & 166.46/119 \\

         \hline 

         S2 & $1.70^{+0.19}_{-0.18}$ & $19.8^{+2.7}_{-2.4}$ & $1.75^{+0.77}_{-0.63}$ & -- & -- & -- & $1.61^{+0.03}_{-0.20}$ & $4.63^{+0.07}_{-0.64}$ & 147.26/118 \\ 

         \hline

         S3 & $1.63^{+0.20}_{-0.21}$ & $17.9^{+2.7}_{-2.4}$ & $1.43^{+0.81}_{-0.79}$ & $6.39^{+0.05}_{-0.05}$ & $<118^*$ & $134^{+81}_{-33}$ & $1.60^{+0.04}_{-0.15}$ & $4.62^{+0.09}_{-0.64}$ & 124.21/116 \\

         \hline 

         \multirow[c]{2}{*}{S4} & $1.59^{+0.21}_{-0.22}$ & $17.6^{+2.7}_{-2.4}$ & $1.56^{+0.92}_{-0.86}$ & $6.39^{+0.05}_{-0.05}$ & $<121^*$ & $107^{+118}_{-6}$ & $1.61^{+0.03}_{-0.22}$ & $4.63^{+0.12}_{-0.67}$ & 122.48/114 \\
         & & & & $\sim4.54$ & $100^*$ & $68^{+12}_{-68}$ & & & \\

         \hline 

         F1 & $1.59^{+0.21}_{-0.20}$ & $17.2^{+2.3}_{-2.0}$ & 1* & $6.39^{+0.05}_{-0.05}$ & $<123^*$ & $146^{+82}_{-31}$ & $1.60^{+0.04}_{-0.16}$ & $4.62^{+0.10}_{-0.48}$ & 125.39/117 \\

         \hline
         
         \multirow[c]{2}{*}{F2} & $1.61^{+0.22}_{-0.20}$ & $17.1^{+2.4}_{-2.0}$ & 1* & $6.39^{+0.05}_{-0.05}$ & $<125^*$ & $131^{+114}_{-7}$ & $1.60^{+0.02}_{-0.20}$ & $4.61^{+0.08}_{-0.47}$ & 122.01/115 \\
         & & & & $\sim7.00$ & $100^*$ & $80^{+38}_{-70}$ & & & \\

         \hline

         \multirow[c]{2}{*}{F3} & $1.61^{+0.22}_{-0.20}$ & $17.4^{+2.5}_{-2.1}$ & 1* & 6.16* & 100* & $37^{+39}_{-37}$ & $1.60^{+0.03}_{-0.18}$ & $4.61^{+0.11}_{-0.64}$ & 139.60/116 \\
         & & & & 6.56* & $290^{+491}_{-226}$ & $205^{+57}_{-183}$ & & & \\

    \end{tabular}
    
    \caption{Best-fit model parameters for the first \textit{XMM}/EPIC observation. Model names starting with ‘S’ refer to those used by \citet{severgnini_swift_2018}, and those starting with ‘F’ refer to those used by \citet{foord_chandra_2025}. The models are further described in the text, along with any fixed parameters and their values. The columns are numbered as follows: (1) Model name (2) Power-law index (3) Column density intrinsic to the source (4) Reflection scaling factor in the \texttt{pexrav} model (5) Rest-frame Gaussian emission line center(s) (6) Gaussian line width(s) (7) Line equivalent width(s). (8) Observed X-ray flux in the 2-10 keV band (9) Rest-frame source luminosity in the 2-10 keV band (10) $\chi^2$ fit statistic and degrees of freedom. Note that values in columns (7), (8), and (9) were calculated from the more sensitive pn detector. An asterisk (*) denotes a fixed parameter. Values preceded by a tilde ($\sim$) indicate poorly-constrained parameters. All quoted uncertainties and limits are at the 90\% level.}

    \label{tab:pn_obs1}
\end{table*}

\subsection{Comparison With Prior Work} \label{subsec:comparison}
Here, we will examine our results side by side with those from \citet{severgnini_swift_2018} and \citet{foord_chandra_2025}. In addition to the double Fe K$\alpha$ line hypothesis, we would like to compare source properties such as the X-ray spectral power-law index, intrinsic source absorption, and the strengths of the reflection component and iron line.

We begin by comparing of the sensitivity of all three instruments. \citet{severgnini_swift_2018} stack 166 ks of \textit{Swift}/XRT data, and their spectrum contains 4300 counts. \textit{Chandra}/ACIS has a lower effective area than \textit{XMM}/EPIC, and \citet{foord_chandra_2025} record 2100 counts in 34 ks. Even with a $30-50\%$ loss of exposure time, each epoch of our data contains $\sim5000$ counts across three detectors. The sensitivity of \textit{XMM} and the combination of these detectors in the EPIC instrument afford us a greater signal-to-noise ratio. Our non-detection of a double iron line in each epoch of our data strengthens the conclusion drawn by \citet{foord_chandra_2025}.

\citet{severgnini_swift_2018} model spectra from MCG+11 in a ``high state", in which the 2-10 keV flux from the source is $\sim4.3 \times 10^{-12}$ erg s$^{-1}$ cm$^{-2}$. This is a factor of 2.7 higher than our flux measurement from the first observation and 2.3 times higher than the second observation. \citet{foord_chandra_2025} observe MCG+11 in somewhat of an intermediate state, with $F_{\text{2-10 keV}} \sim 3.4\times 10^{-12}$ erg s$^{-1}$ cm$^{-2}$. \citet{severgnini_swift_2018} find a 2-10 keV flux ratio between the direct and reflected emission of $\sim0.09$. Our ratio for the first \textit{XMM} observation is 0.11 and 0.09 for the second. These are consistent with the value found in earlier work. As for equivalent widths, those found by \citet{severgnini_swift_2018} and \citet{foord_chandra_2025} agree with our results within the reported 90\% confidence intervals. 

While the absence of a double Fe K$\alpha$ line is common to both our analysis and that of \citet{foord_chandra_2025}, they do find a best-fit energy of 7.56 keV for a second Gaussian line at $2\sigma$ significance. They interpret this as the Fe K$\beta$ line, Ni K$\alpha$ line, or some mixture thereof. Although the addition of this spectral feature does not result in a statistically significant improvement to their fit, they do note that its presence allows for a better constraint on $\Gamma$. Interestingly, we do find a possible emission feature at a similar energy of 7.54 keV (for example, see the fit to model S4 for the second epoch in Table \ref{tab:pn_obs2}). However, our fit is rather insensitive to the line energy and the equivalent width is quite low. Therefore, it is possible that the feature modeled by \citet{foord_chandra_2025} is simply a noise fluctuation.

The fits to stacked \textit{Swift}/XRT data carried out by \citet{severgnini_swift_2018} result in $\Gamma\sim1.5-1.9$. The best-fit model from \citet{foord_chandra_2025} finds $\Gamma\sim1.2-2$. These agree with both epochs of our data ($\Gamma\sim1.3-2.0$), within 90\% confidence intervals. See Figure \ref{fig:gamma_nh_hist} for a visual comparison of these values. In the next section, we will further explore where MCG+11 resides in parameter space among the broader AGN population.

\begin{table*}
    \centering
    \begin{tabular}{c|c|c|c|c|c|c|c|c|c}
    
         \multirow[t]{3}{4em}{(1) Model} & (2) & (3) & (4) & (5) & (6) & (7) & (8) & (9) & (10) \\
         & $\Gamma$ & $N_{\text{H}}$ & $rel_{refl}$ & $E_\text{line}$ & $\sigma_{\text{line}}$ & EW & $F_\text{2-10}$ ($10^{-12}$ & $L_\text{2-10}$ ($10^{42}$ & $\chi^2$/dof \\
         & & ($10^{22}$ cm$^{-2}$) & & (keV) & (eV) & (eV) & erg s$^{-1}$ cm$^{-2}$) & erg s$^{-1}$) \\
         
         \hline

         S1 & $1.27^{+0.17}_{-0.17}$ & $15.6^{+1.5}_{-1.4}$ & -- & -- & -- & -- & $1.87^{+0.05}_{-0.15}$ & $5.36^{+0.11}_{-0.46}$ & 147.02/120 \\

         \hline 

         S2 & $1.54^{+0.20}_{-0.20}$ & $18.9^{+2.7}_{-2.4}$ & $1.48^{+0.75}_{-0.69}$ & -- & -- & -- & $1.86^{+0.04}_{-0.25}$ & $5.33^{+0.12}_{-0.67}$ & 136.30/119 \\ 

         \hline

         S3 & $1.46^{+0.22}_{-0.24}$ & $17.1^{+2.7}_{-2.4}$ & $1.06^{+0.84}_{-0.96}$ & $6.40^{+0.05}_{-0.05}$ & $<86^*$ & $106^{+91}_{-21}$ & $1.86^{+0.04}_{-0.22}$ & $5.34^{+0.10}_{-0.59}$ & 115.52/117 \\

         \hline 

         \multirow[c]{2}{*}{S4} & $1.54^{+0.23}_{-0.25}$ & $17.8^{+2.8}_{-2.6}$ & $1.17^{+0.78}_{-0.86}$ & $6.40^{+0.05}_{-0.05}$ & $100^*$ & $151^{+62}_{-61}$ & $1.85^{+0.03}_{-0.34}$ & $5.32^{+0.16}_{-0.74}$ & 114.11/115 \\
         & & & & $\sim7.54$ & $100^*$ & $43^{+109}_{-43}$ & & & \\

         \hline 

         F1 & $1.45^{+0.21}_{-0.19}$ & $17.0^{+2.3}_{-2.0}$ & 1* & $6.40^{+0.05}_{-0.05}$ & $<88^*$ & $126^{+66}_{-34}$ & $1.86^{+0.04}_{-0.19}$ & $5.34^{+0.10}_{-0.81}$ & 115.66/118 \\

         \hline
         
         \multirow[c]{2}{*}{F2} & $1.45^{+0.20}_{-0.19}$ & $17.0^{+2.2}_{-1.9}$ & 1* & $6.40^{+0.05}_{-0.05}$ & $100^*$ & $103^{+118}_{-21}$ & $1.86^{+0.04}_{-0.20}$ & $5.34^{+0.12}_{-0.65}$ & 116.51/116 \\
         & & & & $\sim6.00$ & $100^*$ & $<55$ & & & \\

         \hline

         \multirow[c]{2}{*}{F3} & $1.47^{+0.21}_{-0.20}$ & $17.3^{+2.4}_{-2.0}$ & 1* & 6.16* & 100* & $42^{+20}_{-42}$ & $1.85^{+0.05}_{-0.19}$ & $5.33^{+0.13}_{-0.47}$ & 139.50/118 \\
         & & & & 6.56* & $100^*$ & $26^{+98}_{-2}$ & & & \\

    \end{tabular}
    
    \caption{Same as Table \ref{tab:pn_obs1} for the second \textit{XMM}/EPIC observation.}

    \label{tab:pn_obs2}
\end{table*}

\section{Discussion} \label{subsec:discussion}

\begin{figure}
    \centering
    \includegraphics[width=\linewidth]{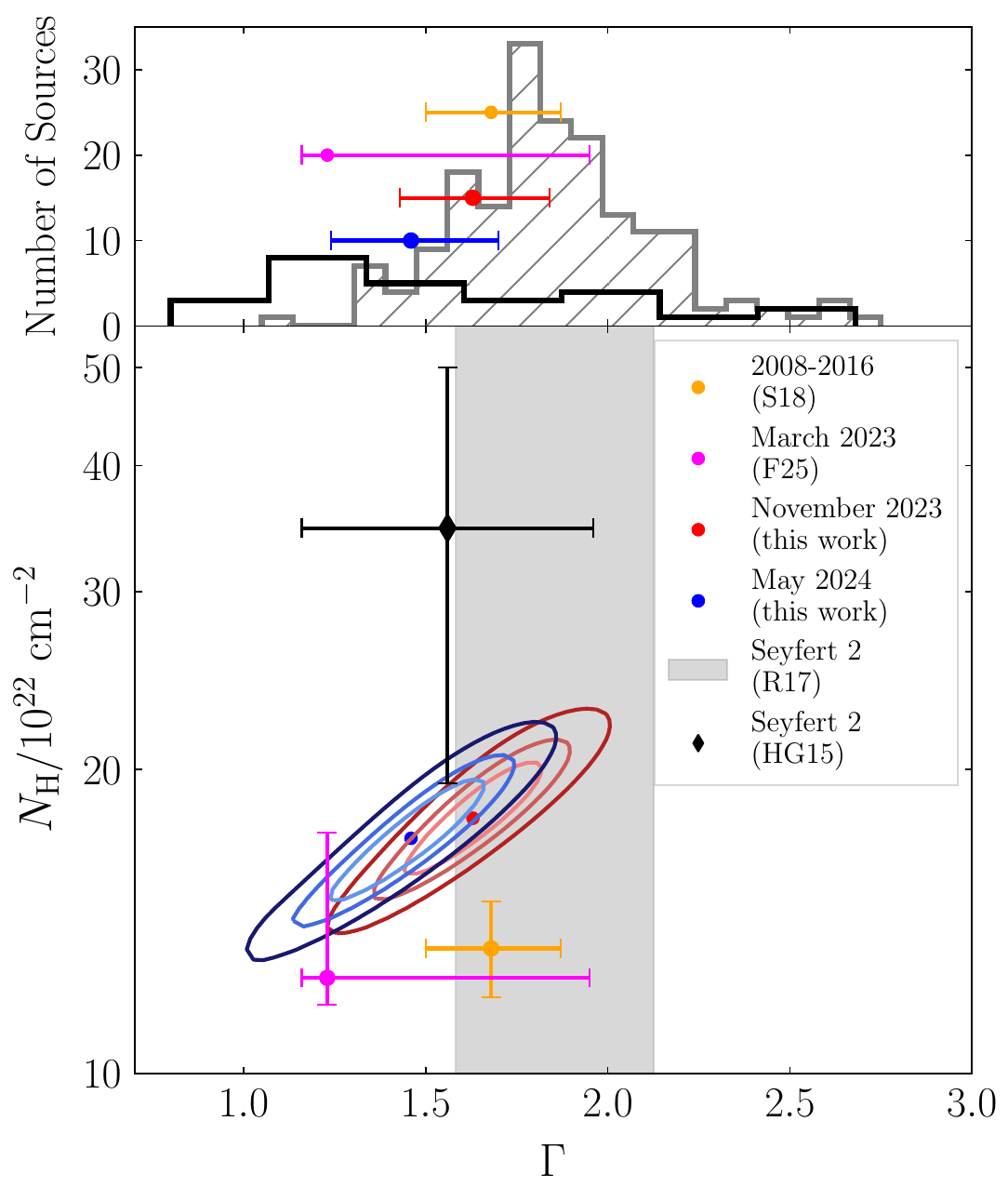}
    \caption{\textit{Top:} In gray (hatched), we show a histogram of a sample of 179 Seyfert 2 photon indices, using the catalog from BASS DR1 \citep{koss_bat_2017} cross-matched with X-ray results presented in \citealt{ricci_bat_2017} (R17). In black, we plot a similar histogram for the 26 Seyfert 2 AGN in \citealt{hernandez-garcia_x-ray_2015} (HG15). Over-plotted on the histograms are the values for MCG+11 obtained in this work, by \citealt{severgnini_swift_2018} (S18), and by \citealt{foord_chandra_2025} (F25). See the bottom panel for the legend. \textit{Bottom:} Column density $N_{\text{H}}$ plotted against photon index $\Gamma$ for MCG+11 as well as summary values for the other Seyfert 2 AGN samples. The gray band from \citet{ricci_bat_2017} spans one standard deviation above and below the mean photon index of the BASS DR1 sample; $N_\text{H}$ values span the range of the plot. The point from \citet{hernandez-garcia_x-ray_2015} represents the mean and standard deviation for $\Gamma$ and $N_{\text{H}}$. All uncertainties from \citet{severgnini_swift_2018} and \citet{foord_chandra_2025} are calculated from 90\% confidence regions.}
    \label{fig:gamma_nh_hist}
\end{figure}

To compare MCG+11 with a larger AGN sample, we used the BAT AGN Spectroscopic Survey Data Release 1 (BASS DR1; \citealt{koss_bat_2017}), selecting only those sources studied with X-ray spectroscopy \citep{ricci_bat_2017} and classified as Seyfert 2 via optical spectroscopy. We find 179 sources that meet these criteria. In the top panel of Figure \ref{fig:gamma_nh_hist}, we plot a gray-hashed histogram of the photon indices from this sample of 179 AGN as measured by \citet{ricci_bat_2017}, for which the mean is $\Gamma_\text{mean, BAT} = 1.85 \pm 0.27$. Also included in the figure are optically-selected Seyfert 2 AGN (\citealt{hernandez-garcia_x-ray_2015}; hereafter HG15) with mean $\Gamma_\text{mean, H-G} = 1.58 \pm 0.46$. The sample presented by HG15 consists of 26 nearby ($z < 0.05$), Seyfert 2 sources with at least 400 counts in multiple epochs with \textit{Chandra} or \textit{XMM-Newton}. The $\Gamma$ measurements corresponding to MCG+11 between 2008 and 2024 are consistent with these distributions at all times.

We can also examine the source's possible variation in $\Gamma$ concurrently with changes in the column density along the line of sight. In the bottom panel of Figure \ref{fig:gamma_nh_hist}, we plot contours for MCG+11 in the $\Gamma-N_{\text{H}}$ plane. The contours were calculated using the \texttt{steppar} command within \textsc{xspec} on a grid of 1600 sample points. We trace $\Delta\chi^2 =$ 2.30, 4.61, and 9.21 contours from the best fit (solid points). We include the HG15 sample and our BASS DR1 sample from the top panel. Note that in the bottom panel, the gray band representing the BASS sample spans the full column density range. This is because Seyfert 2 sources can be classified as either ``obscured" ($10^{22} \text{ cm}^{-2} \leq N_{\text{H}} \leq 10^{24} \text{ cm}^{-2}$) or ``Compton-thick" ($N_{\text{H}} > 10^{24} \text{ cm}^{-2}$) by \citet{ricci_bat_2017}. The derived column densities for MCG+11 appear to be systematically lower than the HG15 AGN, however they are still within 1$\sigma$ of the sample mean. In summary, the $N_\text{H}$ and $\Gamma$ we find for MCG+11 are largely consistent with the Seyfert 2 population.

\section{Conclusions} \label{sec:conclusion}
In this work, we have tested \textit{XMM-Newton}/EPIC observations of the local Seyfert 2 MCG+11--11--032 for the presence of a double Fe K$\alpha$ line, which is a possible signature of a binary supermassive black hole system. Compared with previous work, our approach includes taking two epochs of data spaced 6 months apart to check for signatures of orbital modulation of this emission line. We find that the spectra are best fit with a model that includes Galactic and intrinsic absorption, a power-law X-ray source, slab reflection, and a single Gaussian Fe K$\alpha$ component. The centroid of the Gaussian is consistent with the rest-frame energy of 6.4 keV in both exposures. After comparing with additional models from previous work, we find no evidence of a double line in either epoch of our data. This is in good agreement with \textit{Chandra} data presented by \citet{foord_chandra_2025}. 

We also compared MCG+11 with X-ray and optically-selected samples of local Seyfert 2 galaxies. MCG+11 has a $\Gamma$ and $N_\text{H}$ consistent with these populations. Combined with our spectral fitting, we favor a physical scenario that does not require a binary AGN to explain its X-ray spectral properties. However, while the high signal-to-noise spectroscopy presented here argues against the double iron line interpretation suggested by earlier \textit{Swift}/XRT data, our results do not directly test the binary scenario inferred from the \textit{Swift}/BAT light curve. A combination of multiple techniques is likely required to probe the nature of a SMBHB candidate.

Our study uses CCD detectors, but future studies of SMBHBs will also benefit from high-resolution X-ray spectroscopy enabled by microcalorimetry. This technology relies on converting the energy deposited by a photon into heat, permitting the measurement of X-ray energies to within a few eV. They therefore offer superior spectral resolution to CCDs and can precisely characterize line features. This capability can certainly help with identifying double lines and other spectral features from SMBHBs at high significance.

With its launch in 2023, the \textit{X-ray Imaging and Spectroscopy Mission} (\textit{XRISM}; \citealt{tashiro_status_2020}) has ushered in the era of microcalorimetry with the Resolve instrument \citep{ishisaki_status_2022}. The next decade will see the launch of \textit{NewAthena} and one of its onboard instruments X-IFU (\citealt{cruise_newathena_2025}, \citealt{peille_x-ray_2025}), another microcalorimeter. X-IFU will have an energy resolution of 4 eV at 7 keV as opposed to \textit{XMM}/EPIC, which has a resolution of 150 eV at 6.4 keV. As a result of its large effective area and superior resolution, it will be able to leverage additional theoretical predictions for SMBHB observables. For example, \citet{malewicz_x-ray_2025} produce mock spectra from unequal-mass binaries to explore spectral signatures of preferential accretion. This term refers to the fact that the less massive black hole accretes faster, leading to different ionization states for the two mini-disks. When fitting the X-IFU spectra with a single-AGN model, they find that nearby sources (z = 0.1) with $M_{\text{tot}} = 10^9 M_{\odot}$ show large residuals at soft X-ray energies. These kinds of observations can certainly provide supporting evidence for a SMBHB alongside other methods such as lightcurve variability analysis and a multiwavelength approach.


\begin{acknowledgments}
We thank the anonymous referee for their comments, which greatly improved the manuscript. This work is based on observations obtained with \textit{XMM-Newton}, an ESA science mission with instruments and contributions directly funded by ESA Member States and NASA. Support for this work was provided by NASA through the \textit{XMM-Newton} Guest Observer program (grant number 80NSSC24K0542).
\end{acknowledgments}

\begin{contribution}
Y.A.G. reduced and analyzed the \textit{XMM-Newton} data and drafted the manuscript. T.L. was the PI of the \textit{XMM-Newton} proposal and oversaw the project progress and manuscript writing. R.M. and C.S.R. contributed to the analysis and interpretation. M.K. was a Co-I of the \textit{XMM-Newton} proposal and contributed to the manuscript.


\end{contribution}

%
\facilities{\textit{XMM-Newton}}

\software{\textsc{xspec} \citep{arnaud_xspec_1996}, \textsc{numpy} \citep{harris_array_2020}, \textsc{matplotlib} \citep{hunter_matplotlib_2007}, \textsc{astropy} \citep{astropy_collaboration_astropy_2022}}




\bibliographystyle{aasjournal}
\bibliography{MCG+11-11-032}{}



\end{document}